\documentclass{article}
\usepackage{graphics}
 \usepackage{graphicx}

 \usepackage{epsfig}

\usepackage{amssymb}
\usepackage{amsmath}
 \usepackage{amsthm}

\begin{document}

\title{Study of couplings effect on the performance of a spin-current diode:
Nonequilibrium Green's function based model}

\author{M.˜ Bagheri Tagani,
        H.˜ Rahimpour Soleimani,\\
        \small{Department of physics, University of Guilan, P.O.Box 41335-1914, Rasht, Iran}}

\maketitle

\begin{abstract}
In this paper, spin-dependent transport through a spin diode
composed of a quantum dot coupled to a normal metal and a
ferromagnetic lead is studied. The current polarization and the
spin accumulation are analyzed  using the equations of motion
method within the nonequilibrium Green's function formalism. We
present a suitable method for computing the Green's function
without carrying out any self-consistent calculation. Influence of
coupling strength and magnetic field on the spin current is
studied and  observed that this device can not work as a spin
diode under certain conditions.
\end{abstract}

\section{Introduction}
\label{Introduction} Spin-dependent transport through a quantum
dot (QD) has attracted increasing attention in recent years due
to development in constructing spin-based devices
~\cite{Mikkelsen,Trauzettel,Ping Zhang,Lu,Souza1}. Study of
transport through QDs obtains interesting information about novel
physics phenomena such as Kondo
effect~\cite{Liang,Numata,Weymann1}, spin and Coulomb blockade
effects~\cite{Liu, Romeike, Park, Dong}, spin valve effect
\cite{Konig, Ying}, tunneling magnetoresistance~\cite{Weymann2,
Hamaya}, zero-bias anomaly~\cite{Buizert} and etc. Coupling the QD
to the different contacts can lead to form different devices for
example the spin filters constructed by coupling the QD to the
normal metals~\cite{Engel,Cota} and spin diodes constructed by
attaching the QD to a normal metal and a ferromagnetic
lead~\cite{Merchant}.
\par Recently, spin diodes have been studied both experimentally
and theoretically. C. A. Merchant and N.
Markovi$\acute{c}$~\cite{Merchant} observed diode-like behavior
in a carbon nanotube coupled to a ferromagnetic and a nonmagnetic
lead. The spin diode-like behavior was also reported by Ivan and
co-workers~\cite{Iovan} in resonant tunneling sandwiched by tunnel
barriers with different spin-dependent transparencies. In
addition, spin diode devices were theoretically analyzed in a few
articles. Souza and co-workers~\cite{Souza} studied a
semiconductor QD coupled to a normal metal and a ferromagnetic
metal
 and I. Weymann and co-workers~\cite{Weymann3} studied a spin diode composed of a
 carbon nanotube coupled to a normal metal and a ferromagnetic metal.
  Both of them used rate equations~\cite{Glazman,Mitra} which are valid in the limit $\Gamma_0<<kT$
   that $\Gamma_0$ and $T$ are coupling strength and
  temperature, respectively. Here, we use the nonequilibrium Green's function formalism
  (NEGF)~\cite{Haug} also valid in the limit $\Gamma_0\succeq kT$ to analyze spin diode behavior composed
  of a quantum dot. The Green's function is obtained by means of the
  equations of motion method within the nonequilibrium Green's function
  formalism up to the second order of Hartree-Fock
  approximation~\cite{Swirkowicz}.
  We extract an analytical relation for the electron density without
  performing any self-consistent calculation. The obtained result can be
   useful for studying charge transport through
  mesoscopic systems by means of equations of motion method within Green's function formalism.
  The influence of spin splitting
   on the performance of the device is investigated.
  This splitting can be created by an external magnetic field or
  coupling of the dot to a magnetic substrate. In addition, the
  influence of coupling strength on the diode behavior is analyzed.
Such an examination is impossible by means of the rate equations
because it is just valid in the limit of very weak coupling.

\par The article is organized as follows: the model Hamiltonian and
main formulas are presented in section 2, in section 3 we present
numerical results and discuss the difference between results
obtained from NEGF and rate equations. In the end, some sentences
are given as a conclusion.

 \section{Description of the model}
 \label{Description of the model}

 Hamiltonian describing a quantum dot coupled to a ferromagnetic and a nonmagnetic lead
  is written as
 \begin{align}
H&=\sum_{\alpha,k,\sigma}\varepsilon_{\alpha,k,\sigma}c^{\dag}_{\alpha,k,\sigma}
c_{\alpha,k,\sigma}+\sum_{\sigma}\varepsilon_{\sigma}n_{\sigma}
+Un_{\uparrow}n_{\downarrow}
        \nonumber \label{Eq1}\\
&\quad +
\sum_{\alpha,k,\sigma}[T_{\alpha,k,\sigma}c^{\dag}_{\alpha,k,\sigma}d_{\sigma}+T^{\dag}_{\alpha,k,\sigma}d^{\dag}_{\sigma}c_{\alpha,k,\sigma}]
\end{align}
where $c^{\dag}_{\alpha,k,\sigma}(c_{\alpha,k,\sigma})$ creates
(annihilates) an electron with momentum $k$, spin $\sigma$ in the
lead $\alpha$. $\varepsilon_{\sigma}=\varepsilon_0+\sigma \Delta$
is the energy level of the dot, $\Delta=2g\mu_{B}B$ denotes Zeeman
splitting, $B$ is magnetic field  and $\sigma$ is equal to $1(-1)$
for  $\uparrow(\downarrow)$, respectively.
$d^{\dag}_{\sigma}(d_{\sigma})$ is the creation (annihilation)
operator in the dot and $n_{\sigma}=d^{\dag}_{\sigma}d_{\sigma}$
is the occupation operator. $U$ is on-site Coulomb interaction
strength, $T_{\alpha,k,\sigma}$ describes tunneling between the
dot and the lead $\alpha$ and it is also assumed that the electron
spin is conserved during the tunneling.

\par In order  to analyze  the system,
the nonequilibrium Green's function method has been used. The
retarded Green's function of the dot is
$G^{r}_{\sigma}=-i\Theta(t-t')<\{d_{\sigma}(t),d^{\dag}_{\sigma}(t')\}>$
that in the steady state it only depends on the time difference
$\tau=t-t'$. Hence, it is better to use its Fourier transform
$G^{r}_{\sigma}(\epsilon)=<<d_{\sigma},d^{\dag}_{\sigma}>>_{\epsilon}$.
 Using the equations of motion technique for the nonequilibrium
Green's function up to the second Hartree-Fock approximation, the
spin-dependent Green's function of the dot is given by
\begin{align}
G^{r}_{\sigma}(\epsilon)&=\frac{1-<n_{\bar{\sigma}}>}{\epsilon-\varepsilon_{\sigma}+\frac{i}{2}(\Gamma^{L}_{\sigma}(\epsilon)+\Gamma^{R}_{\sigma}(\epsilon))}+
 \nonumber \label{Eq2}\\
 &\quad
\frac{<n_{\bar{\sigma}}>}{\epsilon-\varepsilon_{\sigma}-U+\frac{i}{2}(\Gamma^{L}_{\sigma}(\epsilon)+\Gamma^{R}_{\sigma}(\epsilon))}
\end{align}
where $\Gamma^{\alpha}_{\sigma}(\epsilon)=2\pi
\sum_{k}|T_{\alpha,k,\sigma}|^2\delta(\epsilon-\varepsilon_{\alpha,k,\sigma})$
is the coupling strength giving rise to the broadening of the dot
levels due to tunneling through the left and right leads and
$\bar{\sigma}$ stands for the opposite spin $\sigma$. In the
following we use the wide band limit i.e. the energy independent
broadening
$\Gamma^{\alpha}_{\sigma}(\epsilon)=\Gamma^{\alpha}_{\sigma}$.
The electron density is given by~\cite{Swirkowicz}

\begin{equation}\label{Eq3}
<n_{\sigma}>=-2\int\frac{d\epsilon}{2\pi}\frac{f^{L}(\epsilon)\Gamma^{L}_{\sigma}+f^{R}(\epsilon)\Gamma^{R}_{\sigma}}{\Gamma^{L}_{\sigma}+\Gamma^{R}_{\sigma}}Im(G^{r}{\sigma}(\epsilon))
\end{equation}
that $f^{\alpha}(\epsilon)$ denotes the Fermi distribution
function of the lead $\alpha$ with the chemical potential
$\mu_\alpha$. Although it seems Eqs.(2,3) should be solved in a
self-consistent manner to obtain the exact Green's function, we
show the Green's function and the electron density  can be
analytically computed. From Eq.\eqref{Eq2} the retarded Green's
function can be written as follows~\cite{Zimbovskaya}
\begin{equation}\label{Eq4}
  G^{r}_{\sigma}(\epsilon)=A_{\sigma}(\epsilon)+B_{\sigma}(\epsilon)<n_{\bar{\sigma}}>
\end{equation}
where
\begin{subequations}
\begin{align}
A_{\sigma}(\epsilon)&=\frac{1}{\epsilon-\varepsilon_{\sigma}+\frac{i}{2}(\Gamma^{L}_{\sigma}+\Gamma^{R}_{\sigma})}\label{Eq5a}\\
B_{\sigma}(\epsilon)&=\frac{U}{(\epsilon-\varepsilon_{\sigma}+\frac{i}{2}(\sum_{\alpha}\Gamma^{\alpha}_{\sigma}))(\epsilon-\varepsilon_{\sigma}-U+\frac{i}{2}(\sum_{\alpha}\Gamma^{\alpha}_{\sigma}))}\label{Eq5b}
\end{align}
\end{subequations}
If Eq.\eqref{Eq4} is substituted into Eq.\eqref{Eq3}, it can be
easily shown that the spin-dependent density is given by
\begin{equation}\label{Eq6}
  <n_{\sigma}>=\frac{Q_{\sigma}+R_{\sigma}Q_{\bar{\sigma}}}{1-R_{\sigma}R_{\bar{\sigma}}}
\end{equation}
where
\begin{subequations}
\begin{align}
Q_{\sigma}&=-2\int\frac{d\epsilon}{2\pi}\frac{\Gamma^{L}_{\sigma} f^{L}(\epsilon)+\Gamma^{R}_{\sigma}f^{R}(\epsilon)}{\Gamma^{L}_{\sigma}+\Gamma^{R}_{\sigma}}Im(A_{\sigma}(\epsilon))   \label{Eq7a}\\
R_{\sigma}&=-2\int\frac{d\epsilon}{2\pi}\frac{\Gamma^{L}_{\sigma}
f^{L}(\epsilon)+\Gamma^{R}_{\sigma}f^{R}(\epsilon)}{\Gamma^{L}_{\sigma}+\Gamma^{R}_{\sigma}}Im(B_{\sigma}(\epsilon))
\label{Eq7b}
\end{align}
\end{subequations}

\par Knowing the electron density Eq.\eqref{Eq6},
the Green's function is obtained from Eq.\eqref{Eq4}. Now, we are
able to compute the current given by ($e=\hbar=1$)~\cite{Jauho}
\begin{equation}\label{Eq8}
  I_{\sigma}=-2\int\frac{d\epsilon}{2\pi}[f^{L}(\epsilon)-f^{R}(\epsilon)]\frac{\Gamma^{L}_{\sigma}\Gamma^{R}_{\sigma}}{\Gamma^{L}_{\sigma}+\Gamma^{R}_{\sigma}}
Im(G^{r}_{\sigma}(\epsilon))
\end{equation}
For simulation purpose, we set $U=4meV$, $kT=212meV$ which is
large enough to guarantee no Kondo effect  and
$\varepsilon_{\sigma}=1meV+\sigma\Delta$ that $\Delta$ is equal
to $1.16\times 10^{-4}meV$ or $0.174 meV$ for $B=10^{-3}T$ and
$1.5 T$, respectively. The coupling strength for the metal lead is
set $\Gamma_{\uparrow}=\Gamma_{\downarrow}=\Gamma_0$ and for
ferromagnetic lead $\Gamma_{\uparrow(\downarrow)}=\Gamma_0(1\pm
 p)$ that $p$ stands for the spin polarization degree
  of the lead. The
chemical potential of the left lead (normal metal) is equal to
zero and for right one, we set $\mu^{R}=-V$, so when the bias is
positive the left lead acts as an emitter and in the negative
bias it acts as a collector.

\section{Simulation results}
\label{simulation results}

\begin{figure}[htb]
\begin{center}
\includegraphics[height=80mm,width=75mm,angle=0]{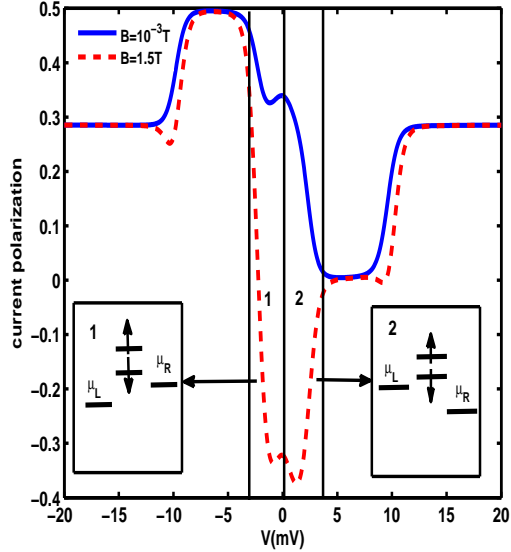}
\caption{ The current polarization as a function of bias in weak
magnetic field (solid) and strong magnetic field (dashed). The
parameters are $p=0.5$ and $\Gamma_0=30\mu eV$. The left and
right insets show the QD levels and the chemical potentials of the
leads in the regions 1 and 2, respectively. Energy difference
between $\uparrow$ and $\downarrow$ levels depends on the
magnitude of the magnetic field. } \label{fig:1}
\end{center}
\end{figure}
Fig. 1 shows the current polarization
$\xi=\frac{I_{\uparrow}-I_{\downarrow}}{I_{\uparrow}+I_{\downarrow}}$
as a function of the bias voltage in different magnetic fields.
It is observed that in positive bias when
$\mu^{R}<\varepsilon_{\sigma}$ and
$\varepsilon_{\sigma}<\mu^{L}<\varepsilon_{\sigma}+U$ (the dot is
singly occupied) the current polarization becomes zero because in
positive bias the normal metal lead acts as an emitter and the
spin-up current is equal to spin-down current due to
$\Gamma^{L}_{\uparrow}=\Gamma^{L}_{\downarrow}$. Note that when
the QD is singly occupied, the current polarization depends on
$(\Gamma_{\uparrow}^{\alpha}-\Gamma_{\downarrow}^{\alpha})$ that
$\alpha$ denotes the emitter lead~\cite{Souza}.
 In negative
bias, the ferromagnetic lead acts as an emitter and the current
polarization becomes maximum when
$\varepsilon_{\sigma}<\mu^{R}<\varepsilon_{\sigma}+U$ because of
$\Gamma^{R}_{\uparrow}>\Gamma^{R}_{\downarrow}$ so that the
spin-up current is bigger than the spin-down current. This
behavior suggests that this system operates as a spin diode in a
definite voltage range. We also observe the current polarization
behaves completely different in the response to weak and strong
magnetic fields in the regions 1 and 2 shown in the fig. 1.
Although our results and the results given in~\cite{Souza} are
nearly the same in weak $B$ and $\Gamma_0$, they are completely
different in strong field because of removing
degeneracy~\cite{Ref28}. The left inset can help us to understand
the current polarization behavior in the region 1 where the
ferromagnetic lead acts as the emitter and the dot levels are
outside the bias window. In the weak magnetic field
($B=10^{-3}T$)  $\varepsilon_\uparrow$ and
$\varepsilon_\downarrow$ are nearly degenerate so that the spin-up
can be occupied faster than the other spin  due to
$\Gamma^{R}_{\uparrow}>\Gamma^{R}_{\downarrow}$ and as a result
$I_{\uparrow}>I_{\downarrow}$, therefore, $\xi$ is positive. In
the strong magnetic field ($B=1.5T$), the levels of the spin-up
and spin-down are split so that the $\varepsilon_{\downarrow}$ is
more energetically accessible hence $I_{\downarrow}>I_{\uparrow}$
and as a result the current polarization becomes negative. The
right inset describes the position of the chemical potentials of
the leads and the energy levels of the dot in the region 2 in
which the normal metal lead acts as the emitter. In weak magnetic
field
 both levels have the same energy the $\uparrow$ electron can
leave the dot faster due to
$\Gamma^{R}_{\uparrow}>\Gamma^{R}_{\downarrow}$ therefore
$I_{\uparrow}$ is bigger than $I_{\downarrow}$ and the current
polarization is positive, But in the strong magnetic field the
$\downarrow$ electron state is occupied faster because it is more
accessible and as a consequence, $I_{\downarrow}>I_{\uparrow}$ so
that the current polarization becomes negative. When
$|V|>\varepsilon_{\sigma}+U$ that the dot is doubly occupied the
current polarization will be constant due to interplay between
the spin accumulation and the electron-electron interaction.
\begin{figure}[htb]
\begin{center}
\includegraphics[height=80mm,width=75mm,angle=0]{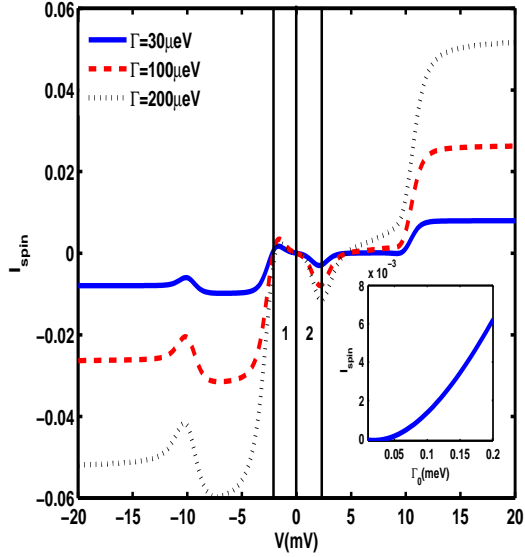}
\caption{ The spin current versus bias in different $\Gamma_0$. We
set $B=1.5T$. Inset shows variation of the spin current as
function of broadening in $V=9mV$. } \label{fig:2}
\end{center}
\end{figure}
\par
Fig. 2 plots the spin current
$I_{spin}=I_{\uparrow}-I_{\downarrow}$ versus the bias voltage for
different coupling strengths. As we expect with increase of
$\Gamma_0$ the current is enhanced because of faster tunneling.
We observe when
$|\varepsilon_{\sigma}|<|eV|<|\varepsilon_{\sigma}+U|$, the spin
current as well as the current polarization becomes zero just in
the limit of low or intermediate $\Gamma_0$ and with increasing
the coupling strength the current polarization deviates from zero
and as a result the device can not work as a spin diode. The
dependence of the spin current on $\Gamma_0$ is shown in the inset
in the conditions that the dot is singly occupied. It is observed
that NEGF and rate equation  obtain the same result for
$\Gamma_0<0.05meV$, but the latter one could not describe the
system well for $\Gamma_0>0.05meV$. Indeed, the rate equation
predicts $\xi=0$ for any value of $\Gamma_0$. The unexpected
positive and negative spin current has been also observed in the
regions 1 and 2, respectively. Note that the current is measured
from the left lead i.e. the current is positive in the positive
bias. The positive spin current in the region 1 means that
$I_{\downarrow}$ is bigger than $I_{\uparrow}$ because of
$\varepsilon_{\downarrow}<\varepsilon_{\uparrow}$. The negative
spin current in the region 2 shows that
$I_{\downarrow}>I_{\uparrow}$ due to the same reason.
\begin{figure}[htb]
\begin{center}
\includegraphics[height=80mm,width=100mm,angle=0]{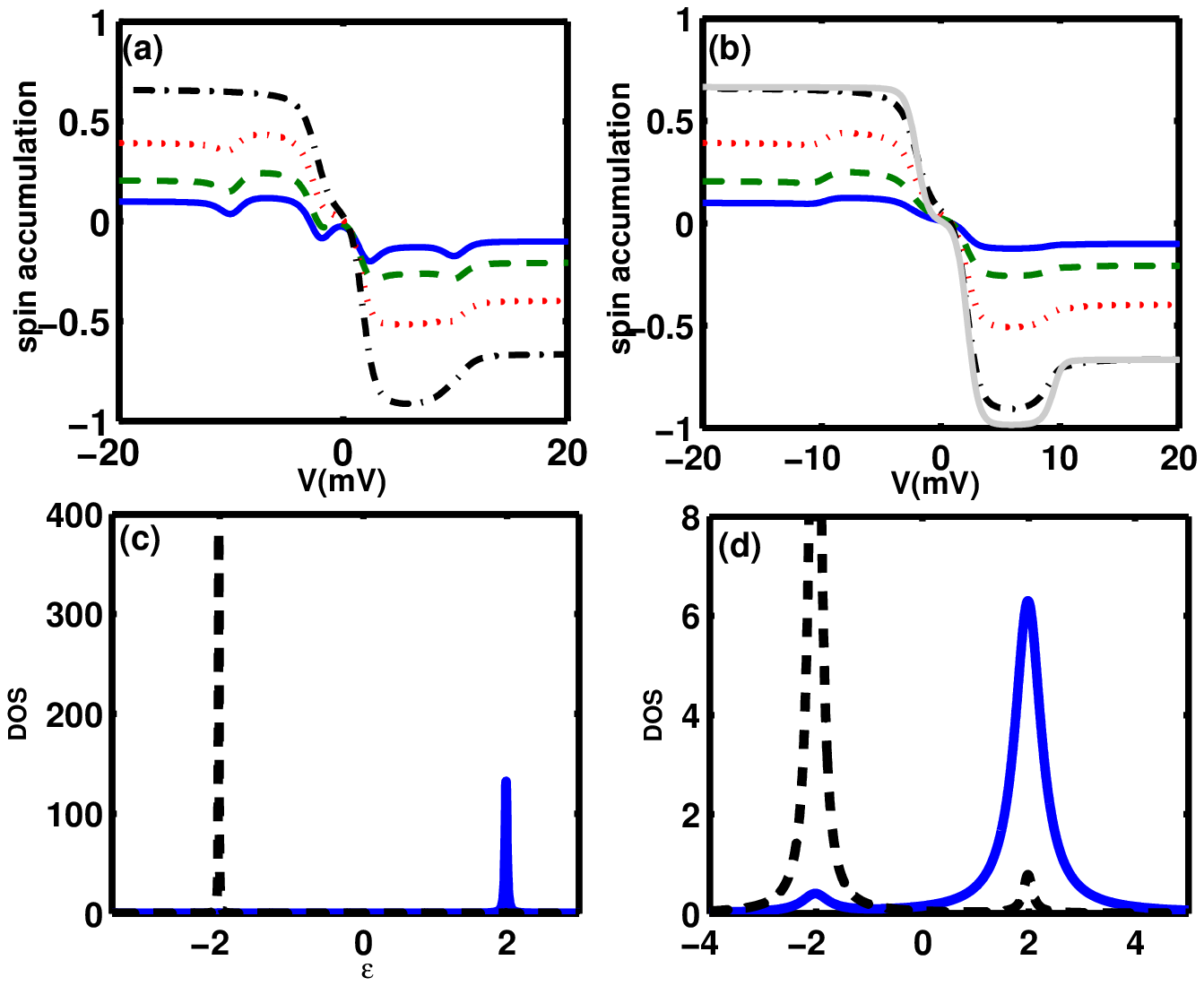}
\caption{ The spin accumulation as a function of polarization in
(a) strong magnetic field and (b) weak magnetic field.We set
$p=0.2$(solid), $p=0.4$ (dashed), $p=0.7$ (dotted) and $p=1$
(dash-dotted). $\Gamma_0=200\mu eV$ in all plots except the gray
line in (b) equal to 10$\mu eV$. (c) and (d) show DOS  (spin-down
(dashed) and spin-up (solid)) in weak and strong couplings,
respectively.} \label{fig:3}
\end{center}
\end{figure}
\par  Fig. 3 shows the spin accumulation $m=n_{\uparrow}-n_{\downarrow}$ as a function of the
bias voltage. It is observed that in the negative bias the spin
accumulation is positive because in this situation the right lead
acts as the emitter and the $\uparrow$ electron is injected faster
and hence the spin-up population is more. In negative bias that
the right lead acts as the collector, the spin accumulation is
negative because a $\downarrow$ electron has to stay inside the
dot for a longer time. Such a behavior was recently reported
using the rate equations method. Here, there is a significant
difference between the results presented in the Ref~\cite{Souza}
and our results obtained using the Green's function method. In
$p=1$ that the right lead is a half metal and in the positive
bias when the energy level of the dot is inside the bias window,
the spin accumulation is equal to -1 (the gray line in the fig.
3b) in the weak broadening ($\Gamma_0=10\mu eV$). Indeed, the
$\downarrow$ electron injected from the left lead can not leave
the dot and since the second level $\varepsilon_{\sigma}+U$ is
outside the bias window, the dot contains only a $\downarrow$
electron. But in the other case $\Gamma_0=200 \mu eV$, we observe
the spin accumulation is about -0.8. This difference originates
due to
 different density of states (DOS) in these
couplings. In the weak coupling (fig. 3c) the DOS of the
$\uparrow$ electron has a peak outside the bias window but the
peak of the DOS of the $\downarrow$ electron is inside the bias
window and as a result the $\downarrow$ electron only exists
inside the dot. In the strong coupling (fig. 3d) the DOS is
 broadened and the DOS of the $\uparrow$ electron has a peak inside
the bias window therefore the fraction of the $\uparrow$ electron
can exist inside the dot leading to the decrease of the spin
accumulation.

\section{Conclusion}
\label{conclusion} In this paper, spin-dependent transport
through a  quantum dot attached to a normal metal and a
ferromagnetic lead is studied by using the nonequilibrium Green's
function formalism. We examine under what conditions this device
can work as a spin diode. We also investigate the influence of
magnetic field on the current polarization. It is observed that
coupling strength has an important role in the performance of the
device. More specifically, in the strong coupling  this device
can not
 work properly. It is also observed when the dot is
coupled to a half metal lead,  the spin accumulation is equal to
-1 in the weak coupling but it deviates from -1 in the strong
coupling.

\bibliographystyle{model1a-num-names}
\bibliography{<your-bib-database>}

\begin{thebibliography}{00}

\bibitem{Mikkelsen}
 M.~H.~Mikkelsen, J.~Berezovsky, N.~G.~Stoltz,
 L.~A.~Coldren, D.~D.~Awschalom,  Nature Physics 3 (2007) 770 - 773.

 \bibitem{Trauzettel}
 B.~Trauzettel, D.~V.~Bulaev, D.~Loss,
 G.~Burkard,
 Nature Physics 3 (2007) 192 - 196.

\bibitem{Ping Zhang}
P.~Zhang,~Qi-Kun.~Xue~,~X.~C.~Xie, Phys. Rev. Lett 91 (2003)
196602.


 \bibitem{Lu}
 H.~Z.~Lu, S.~Q.~Shen,
 Phys. Rev. B 77 (2008) 235309.


 \bibitem{Souza1}
 F.~M.~Souza, A.~P.~Jauho, J.~C.~Egues,
 Phys. Rev. B 78 (2008) 155303.

\bibitem{Liang}
W.~Liang, M.~P.~Shores, M.~Bockrath, J.~R.~Long, H.~Park, Nature
417 (2002) 725-729.

\bibitem{Numata}
Takahide Numata, Yunori Nisikawa, Akira Oguri, Alex C. Hewson,
 Phys. Rev. B 80 (2009) 155330.

\bibitem{Weymann1}
I.~Weymann, J.~Barna$\acute{s}$, Phys. Rev. B 81 (2010) 035331.

\bibitem{Liu}
H. W. Liu, T. Fujisawa, Y. Ono, H. Inokawa, A. Fujiwara, K.
Takashina, Y. Hirayama,  Phys. Rev. B 77 (2008) 073310.

\bibitem{Romeike}
C. Romeike, M. R. Wegewijs, M. Ruben, W. Wenzel, H. Schoeller,
Phys. Rev. B 75 (2007) 064404.

\bibitem{Park}
J. Park, A. N. Pasupathy, J. I. Goldsmith, C. Chang, Y. Yaish, J.
R. Petta, M. Rinkoski, J. P. Sethna, H. D. Abru$\tilde{n}$a, P. L.
McEuen, D. C. Ralph,  Nature 417 (2002) 722-725.

\bibitem{Dong}
B. Dong, H. L. Cui, X. L. Lei,  Phys. Rev. B 69 (2004) 035324.

\bibitem{Konig}
J. K$\ddot{o}$nig, J. Martinek,  Phys. Rev. Lett. 90 (2003)
166602.

\bibitem{Ying}
Y. Ying, G. Jin,  Phys. Lett. A 374 (2010) 3758-3761.

\bibitem{Weymann2}
I. Weymann,  Phys. Rev. B 75 (2007) 195339.

\bibitem{Hamaya}
K. Hamaya, S. Masubuchi, M. Kawamura, T. Machida, M. Jung, K.
Shibata, K. Hirakawa, T. Taniyama, S. Ishida, Y. Arakawa, J.
Appl. Phys. Lett 90 (2007) 053108.

\bibitem{Buizert}
C. Buizert, A. Oiwa, K. Shibata, K. Hirakawa, S. Tarucha, Phys.
Rev. Lett 99 (2007) 136806.


\bibitem{Engel}
H. A. Engel, D. Loss,  Phys. Rev. B 65 (2002) 195321.

\bibitem{Cota}
E. Cota, R. Aguado, G. Platero, Phys. Rev. Lett. 94 (2005) 107202.


\bibitem{Merchant}
C. A. Merchant, N. Markovi$\acute{c}$, J. Appl. Phys. 105 (2009)
07c711-14.

\bibitem{Iovan}
A. Iovan, S. Andersson, Yu. G. Naidyuk, A. Vedyaev, B. Dieny, V.
Korenivski, Nano Lett. 8(3) (2008) 805-809.

\bibitem{Souza}
F. M. Souza, J. C. Egues, A. P. Jauho, Phys. Rev. B 75 (2007)
165303.

\bibitem{Weymann3}
I. Weymann, J. Barna´s,  J. Appl. Phys. Lett. 92 (2008) 103127.



\bibitem{Glazman}
L. I. Glazman, K. A. Matveev,  Pi$s^{'}$ma Zh. $\acute{E}$ksp.
Teor. Fiz. 48 (1988) 403.


\bibitem{Mitra} A.~Mitra, I.~Aleiner, A.J.~Millis,
Phys. Rev. B 69 (2004) 245302.

\bibitem{Haug}
H. Haug and A. P. Jauho, Quantum Kinetics in Transport and Optics
of Semiconductors (Springer, Heidelberg)1996.

\bibitem{Swirkowicz}
R. $\acute{S}$wirkowicz, J. Barna$\acute{s}$, M.
Wilczy$\acute{n}$ski,  J. Phys.: Condens. Matter 14 (2002) 2011.

\bibitem{Zimbovskaya}
N.~A.~Zimbovskaya,  Phys. Rev. B 78 (2008) 035331.

\bibitem{Jauho}
A. P. Jauho, N. S. Wingreen, Y. Meir, Phys. Rev. B 50 (1994) 5528.

\bibitem{Ref28}
In the Ref.~\cite{Souza}, no magnetic field was applied and hence
energy levels were degenerate.

 \end{thebibliography}

\end{document}